\documentclass[10pt, conference]{IEEEtran}

\newif\ifSPACEHACK
\SPACEHACKtrue 

\newif\ifDEBUG
\DEBUGtrue

\newif\ifANONYMOUS
\ANONYMOUSfalse

\usepackage{amsmath,amssymb,amsfonts}
\usepackage{algorithmic}
\usepackage{graphicx}
\usepackage{textcomp}
\usepackage{xcolor}
\usepackage{booktabs} 
\usepackage{xspace} 
\usepackage[normalem]{ulem}
\usepackage{makecell}
\usepackage{tcolorbox}
\usepackage{enumitem}
\usepackage{siunitx}
\usepackage[vskip=1em,font=itshape,leftmargin=2em,rightmargin=2em]{quoting}
\usepackage{hyperref}

\usepackage[noadjust]{cite}

\usepackage{soul}
\usepackage{multirow}
\usepackage{array,booktabs}
\newcolumntype{M}[1]{>{\centering\arraybackslash}m{#1}}
\usepackage{amsfonts}
\usepackage{amsmath}
\usepackage{wrapfig, lipsum}
\usepackage{subfigure}

\ifDEBUG
    \newcommand{\JD}[1]{\textcolor{purple}{[JD:#1]}}
    \newcommand{\AG}[1]{\textcolor{olive}{[AG:#1]}}
    \newcommand{\WJ}[1]{\textcolor{olive}{[WJ:#1]}}
    \newcommand{\GKT}[1]{\textcolor{brown}{[GKT:#1]}}
    \newcommand{\MH}[1]{\textcolor{pink}{[MH:#1]}}
    \newcommand{\NMS}[1]{\textcolor{orange}{[NMS:#1]}}
    \newcommand{\WPM}[1]{\textcolor{red}{[Trey: #1]}}
    \newcommand{\TRS}[1]{\textcolor{teal}{[Taylor: #1]}}

    \newcommand{\TODO}[1]{\hl{#1}}
\else
    \newcommand{\JD}[1]{}
    \newcommand{\AG}[1]{}
    \newcommand{\WJ}[1]{}
    \newcommand{\GKT}[1]{}
    \newcommand{\NS}[1]{}
    \newcommand{\NV}[1]{}
    \newcommand{\NMS}[1]{}
    \newcommand{\WPM}[1]{}
    \newcommand{\TRS}[1]{}
    \newcommand{\MH}[1]{}
    
    \newcommand{\TODO}[1]{}
\fi

\newif\ifSPACEHACK
\SPACEHACKfalse

\newcommand{\myparagraph}[1]{\vspace{0.07cm}\noindent\ul{\emph{\textbf{#1}}}\noindent{}}

\newcommand{\new}[1]{\textcolor{black}{#1}}

\ifSPACEHACK
    \titlespacing*\section{0pt}{4pt plus 2pt minus 2pt}{4pt plus 2pt minus 2pt}
    \titlespacing*\subsection{0pt}{4pt plus 1.5pt minus 1.5pt}{4pt plus 1.5pt minus 1.5pt}
    \titlespacing*\subsubsection{0pt}{3pt plus 1.5pt minus 1.5pt}{3pt plus 1.5pt minus 1.5pt}
    \titlespacing*\paragraph{0pt}{2pt plus 1.5pt minus 1.5pt}{2pt plus 1.5pt minus 1.5pt}
\fi
\usepackage{balance} 

\usepackage{amsmath}
\usepackage{cleveref}

\crefformat{section}{\S#2#1#3}
\crefname{figure}{Figure}{Figures}
\crefname{appendix}{Appendix}{Appendices}
\crefname{table}{Table}{Tables}
\crefname{algorithm}{Algorithm}{Algorithms}
\crefname{listing}{Listing}{Listings}
\crefname{theorem}{Theorem}{Theorems}
\crefname{thm}{Theorem}{Theorems}
\crefname{lemma}{Lemma}{Lemmata}
\crefname{equation}{Eqt.}{Eqts.}
\crefformat{Grammar}{Grammar #1}

\usepackage{enumitem}
\usepackage{booktabs}
\usepackage{svg}
\usepackage{listings}

\newcommand{\ie}{\textit{i.e.,} }
\newcommand{\eg}{\textit{e.g.,} }
\newcommand{\etal}{\textit{et al.}\xspace}

\newcommand{\code}[1]{{\small\texttt{#1}}\xspace}

\newcommand{\HF}{{Hugging Face}\xspace}
\newcommand{\HFs}{{Hugging Face's}\xspace}

\newcommand{\inlinequote}[1]{``\emph{#1}''}

\begin{document}

\newcommand{\interviewNum}{12\xspace}
\newcommand{\interviewNumPRO}{9\xspace}
\newcommand{\interviewRate}{24\%\xspace}

\newcommand{\PTMWDatasets}{{12,811}\xspace}
\newcommand{\PTMWUsableDatasets}{{7,529}\xspace}
\newcommand{\PTMWOUsableDatasets}{{5,282}\xspace}
\newcommand{\PTMWUsableDatasetPercentage}{{58\%}\xspace}
\newcommand{\NUsablePTMDatasets}{{10}\xspace}

\newcommand{\HFNumUsers}{{18,348}\xspace}
\newcommand{\HFNumPROUsers}{{690}\xspace}
\newcommand{\HFNumRegularUsers}{{17,658}\xspace}

\newcommand{\PTMDatasetNPackages}{{63,182}\xspace}
\newcommand{\PTMDatasetPercentage}{{99.7\%}\xspace}
\newcommand{\PTMDatasetFailedPackages}{{186}\xspace}
\newcommand{\PTMDatasetFailedPercentage}{{0.3\%}\xspace}

\newcommand{\PTMDatasetNReposWithSignedCommits}{{132}\xspace}
\newcommand{\PTMDatasetPercentOfSignedCommits}{{0.21\%}\xspace}

\newcommand{\PercentOfVerifiedOrgs}{{3.19\%}\xspace}
\newcommand{\NOrganizations}{{6,243}\xspace}
\newcommand{\NVerifedOrgs}{{199}\xspace}

\newcommand{\NOfRepositoriesWithMalware}{{1}\xspace}
\newcommand{\PercentageOfRepositoriesWithMalware}{{0.002\%}\xspace}
\newcommand{\TotalRepositoriesForMalwareScanning}{{63,366}\xspace}


\newcommand{\mailCandidate}{1331\xspace}
\newcommand{\interviewCandidate}{100\xspace}
\newcommand{\surveyCandidate}{XX\xspace}




\ifANONYMOUS
    \author{Anonymous author(s)}
\else
    \author{
    \IEEEauthorblockN{Wenxin Jiang\IEEEauthorrefmark{1}, Nicholas Synovic\IEEEauthorrefmark{2}, Matt Hyatt\IEEEauthorrefmark{2}, Taylor R. Schorlemmer\IEEEauthorrefmark{1}, Rohan Sethi\IEEEauthorrefmark{2},\\ Yung-Hsiang Lu\IEEEauthorrefmark{1}, George K. Thiruvathukal\IEEEauthorrefmark{2}, James C. Davis\IEEEauthorrefmark{1}}
    \IEEEauthorblockA{\IEEEauthorrefmark{1}Purdue University and \IEEEauthorrefmark{2}Loyola University Chicago}
}



\title{An Empirical Study of Pre-Trained Model Reuse in the Hugging Face Deep Learning Model Registry}
\maketitle


    



    



\fi

\begin{abstract} \label{sec: abstract}
Deep Neural Networks (DNNs) are being adopted as components in software systems. 
Creating and specializing DNNs from scratch has grown increasingly difficult as state-of-the-art architectures grow more complex.
Following the path of traditional software engineering, machine learning engineers have begun to reuse large-scale pre-trained models (PTMs) and fine-tune these models for downstream tasks.
Prior works have studied reuse practices for traditional software packages to guide software engineers towards better package maintenance and dependency management.
We lack a similar foundation of knowledge to guide behaviors in pre-trained model ecosystems. 

In this work, we present the first empirical investigation of PTM reuse.
We interviewed \interviewNum practitioners from the most popular PTM ecosystem, \HF, to learn the practices and challenges of PTM reuse.
From this data, we model the decision-making process for PTM reuse.
Based on the identified practices, we describe useful attributes for model reuse, including provenance, reproducibility, and portability.
Three challenges for PTM reuse are missing attributes, discrepancies between claimed and actual performance, and model risks.
We substantiate these identified challenges with systematic measurements in the \HF ecosystem.
Our work informs future directions on optimizing deep learning ecosystems by automated measuring useful attributes and potential attacks, and envision future research on infrastructure and standardization for model registries.  
\end{abstract}

\begin{IEEEkeywords}
Software reuse,
Empirical software engineering,
Machine learning,
Deep learning,
Software supply chain,
Engineering decision making,
Cybersecurity,
Trust
\end{IEEEkeywords}


\section{Introduction} \label{Intro}

Package reuse has transformed software engineering in programming languages such as JavaScript and Python~\cite{raymond1999cathedral, abdalkareem2017developers}, 
and is transforming deep learning model engineering~\new{\cite{Gopalakrishna2022IoTPractices}}.
Deep Neural Networks (DNNs) are widely used in modern software systems, such as image recognition in autonomous vehicles~\cite{Garcia2020AVBugsSHORT}.
Engineering a DNN is challenging due to the capital and operating expenses of training models~\cite{patterson2021carbon}
and variation in deep learning libraries~\cite{Pham2020AnalysisofVarianceinDLSWSystems}.
These problems can be addressed by reusing \emph{pre-trained DNN models} (PTMs) to amortize DNN development costs across multiple projects and organizations~\cite{Han2021PTM}.
PTMs are shared via \emph{Deep Learning (DL) model registries}, which are modeled on traditional software package registries such as NPM and provide packages with model architectures, weights, licenses, \new{and other metadata}.
DL model registries enable reuse-driven DNN engineering~\cite{TensorFlowHubIntroduction,HuggingFacePaper2020}
As~\cref{fig:downloadrate} shows, PTM reuse is now appreciable~\cite{Han2021PTM}.

{
\begin{figure}[t]
    \centering
    \includegraphics[width=0.95\linewidth]{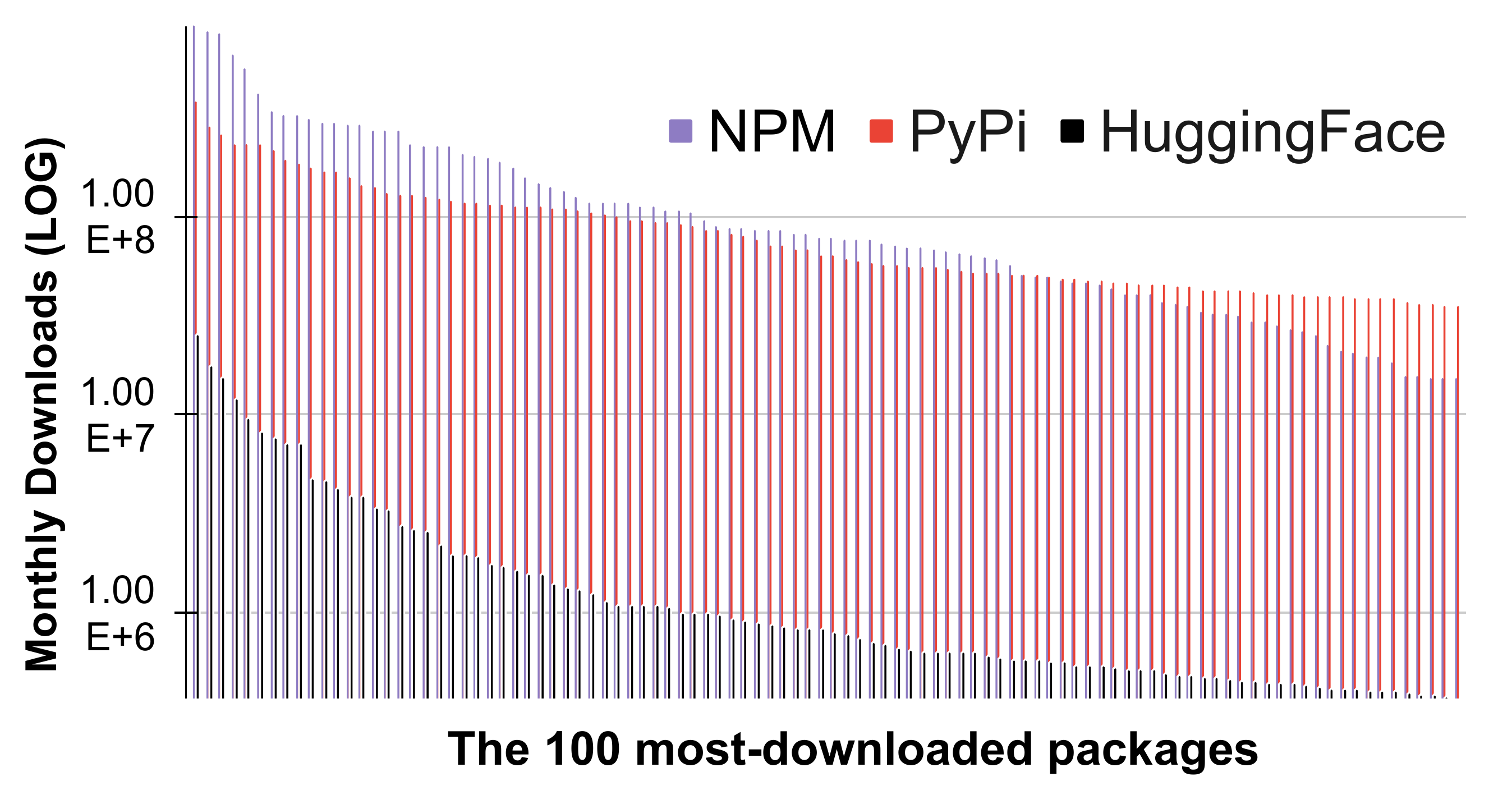}
    \caption{
    Package download rates in two software package registries, NPM and PyPi, and the leading DL model registry, \HF.
    Many \HF model packages have high download rates, though with a rapid drop-off.
    }
    \label{fig:downloadrate}
\end{figure}
}

Reusability and trustworthiness problems in software package registries impact the relevant ecosystems 
~\cite{Zimmermann2019SecurityThreatsinNPMEcosystem, Zahan2022WeakLinksinNPMSupplyChain, zerouali2022impact}.
Much is known about the practices and challenges of reusing traditional software packages~\cite{Jadhav2009EvaluatingSelectingSWPackages, Decan2019DependencyNetworkEvolution,okafor2022sok}, but how this knowledge transfers to the reuse of PTM packages has not been investigated.
Existing software engineering knowledge describes how large companies manage private models~\cite{Amazon2018MLModelManagementChallenge, Amershi2019SE4MLCaseStudy}.
However, we do not know how small-to-large engineering teams reuse models in DL model registries nor what challenges they experience.


%



In this work, we present the first empirical study of pre-trained model reuse.
We took a mixed-methods approach
to identify diverse phenomena for future investigation~\cite{MixedMethodsResearch}.
We focused our study on the \HF DL model registry, which is the largest PTM registry at present~\cite{2022JiangEmpirical}.
First, we interviewed \interviewNum~\HF practitioners to understand the practices and challenges of PTM reuse.
Second, \new{we complemented this qualitative data with measurements of the \HF registry}.
We built a dataflow model of the creation and distribution process for PTMs in \HF and collected and analyzed a dataset of \PTMDatasetNPackages PTM packages.


Our findings indicate that PTM reuse workflows are similar to those for traditional software package reuse, but that engineers follow practices and experience challenges specific to deep learning.
Based on our interview data, we: present the first decision-making workflow for PTM reuse, identify useful PTM attributes and three common challenges, and discuss the extent to which existing techniques meet these challenges (\cref{sec:RQ1}---\cref{sec:RQ3}).
Our dataflow model of PTM distribution found several vectors for software supply chain attacks (\cref{sec:RQ4}).
Our analysis 
shows unique properties of the PTM package ecosystem relative to traditional software package ecosystems, notably that the attributes and decision-making workflow are more complex (\cref{sec:discussion}).
We share the \emph{HFTorrent} dataset of \PTMDatasetNPackages PTM package histories for further analysis (\cref{sec:HFTorrent}).
We conclude by discussing four new research problems for further study (\cref{sec:discussion}).
\ul{Our contributions are}:
\begin{itemize}
    \item We depict a decision-making workflow for PTM reuse, and identify three challenges for PTM reuse
    (\cref{sec:RQ1}---\cref{sec:RQ3}).
    \item We measure the risks of collaboration in \HF.
    We identify several potential software supply chain concerns facing PTM reusers (\cref{sec:RQ4}).
    \item We publish the \emph{HFTorrent} dataset of \PTMDatasetNPackages PTM packages for future analysis (\cref{sec:HFTorrent}).
    \item We identified unique properties of PTM package reuse to guide future research on model audit, infrastructure, standardization, and attack detection (\cref{sec:discussion}).
    
\end{itemize}

\noindent\textbf{\underline{Significance}}:
PTM reuse reduces the engineering costs of employing DNNs in industry.
This paper describes the first investigation of PTM reuse from a software engineering perspective.
We \new{are the first to}
  (1) capture the decision-making workflow and challenges for PTM reuse;
  (2) determine attributes of PTMs that facilitate reuse;
  and
  (3) measure risks of PTM reuse in the \HF DL model registry.
Our findings can help PTM maintainers and registries improve the quality of their offerings, and show opportunities for software engineering tools to support PTM reusers in this process.

\section{Background and Related Work} 
\label{sec:Background}


\subsection{Software Package Reuse}
\label{sec:SoftwarePackageReuse}

\begin{figure}[t]
    \centering
    \includegraphics[width=0.80\linewidth]{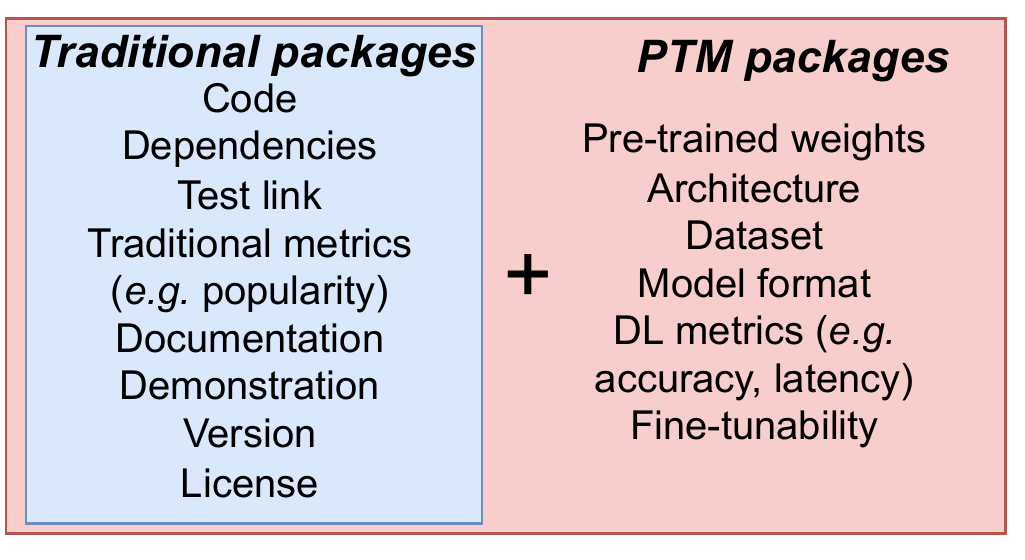}
    \caption{
    Components of traditional packages~\cite{registries_nine_2020-2} and PTM packages~\cite{Neptune2022MLModelRegistry}.
    A PTM package includes all the components of a traditional package, plus some DL-specific parts.
    }
    \label{fig:packages}
\end{figure}

Software package registries store versioned packaged software, associated metadata, documentation, and configurations~\cite{registries_nine_2020-2}.
Similarly, deep learning (DL) model registries distribute PTMs with metadata, a model card (\ie documentation), relevant configurations, and versions of pre-trained weights~\cite{Neptune2022MLModelRegistry}.
DL model registries are an important component of the DL ecosystem~\cite{Smith2020HarnessingMLEcosystem}.
As shown in \cref{fig:packages}, PTM packages may contain more component than traditional packages, including weights, datasets, and performance metrics.

Evaluating and selecting software packages is a difficult, but essential, activity for package reuse~\cite{Jadhav2009EvaluatingSelectingSWPackages}.
Prior work shows that engineers may improve their software selection with insights into the decision-making process and an understanding of relevant factors~\cite{Jadhav2009EvaluatingSelectingSWPackages, Jadhav2011Framework4EvaluationaandSelectionofSWPackages, Bianco2011SurveyonOpenSourceSWTrustworthiness}.
Existing literature focuses on practices in traditional software package registries, such as NPM~\cite{Zerouali2019DiversityofSWPackagePopularityMetrics} 
and Maven~\cite{Mitropoulos2014BugCataLogofMaven, SotoValero2019EmergenceofSWDiversityinMavenCentral}.
The extent to which reuse practices for traditional software will transfer to the reuse of PTM packages is unclear.


Reproducibility
is another important aspect of software package reuse~\cite{Pineau2020}.
In traditional software packages,
  Goswami \etal found that 38\% of explored NPM package versions are non-reproducible~\cite{Goswami2020ReproducibilityofNPMPackages}.
Similarly, Vu \etal 
highlighted existing discrepancies at different levels of granularity in PyPi~\cite{Vu2021LastPyMileSHORT}.
Following the machine learning scientific research community~\cite{Huston2018AIfacesReproducibilityCrisis},
the software engineering community has just begun to study concerns in DL model registries~\cite{Jiang2022PTMSupplyChain}. 
We offer an early software engineering view on this topic.

\subsection{Pre-Trained Model Reuse}
\label{sec:PreTrainedModelReuse}



PTM reuse is necessitated by the emergence of large-scale models, and is enabled by learning and compression techniques, including \emph{transfer learning}~\cite{Pan2010TransferLearning}, 
\emph{quantization and pruning}~\cite{liang_pruning_2021},
\emph{knowledge distillation}~\cite{hinton_distilling_2015}, and \emph{data labeling}~\cite{Dube2019DataLabeling4TransferLearning}.
Through transfer learning, DNNs can be pre-trained on large datasets and fine-tuned to solve specialized tasks, leveraging a PTM's knowledge of one task to better teach it a similar task~\cite{Pan2010TransferLearning,zhuang_comprehensive_2020}. 
Using quantization and pruning methods, PTMs can be optimized for latency- or energy-sensitive contexts, such as on edge devices, without compromising accuracy~\cite{liang_pruning_2021}.
Via knowledge distillation, PTMs can be used to teach a smaller model, yielding good performance and reduced computational costs~\cite{hinton_distilling_2015}.
Engineers can also use PTMs to automatically label datasets~\cite{Dube2019DataLabeling4TransferLearning}.




Practitioners from major technology companies report challenges in model management and model reliability~\cite{Amershi2019SE4MLCaseStudy, Amazon2018MLModelManagementChallenge}.
Schelter \etal summarized model validation challenges, including decisions on model retraining, metadata querying, and adversarial settings~\cite{Amazon2018MLModelManagementChallenge}.
Rahman \etal highlighted that the behavior of ML models can be easily affected because of their data-driven nature~\cite{Rahman2019MLSEinPractice}. 
To better reuse the PTMs, it is important to monitor the performance of deployed models, track changes in data characteristics, and to retrain and re-validate them frequently.

One way to address the management problems is to use a \textbf{DL model registry}, which is defined as: \emph{a collaborative model hub where teams can share DL models}~\cite{Databricks2022ModelRegistry, Neptune2022MLModelRegistry}.
The DL model registry concept imitates traditional software package registries such as NPM~\cite{NPM} and PyPi~\cite{PyPi}.
Through web searches, we identified several prominent DL model registries, including \HF~\cite{HuggingFaceWeb1}, TensorFlow Hub~\cite{tfhub}, PyTorch Hub~\cite{PytorchHub}, and ONNX Model Zoo~\cite{ONNXModelZoo}.
Among all registries we examined~\cite{HuggingFaceWeb1, tfhub, ONNXModelZoo, ModelZooWeb, NVIDIANGC, PytorchHub, ModelhubWeb1}, \HF offers the largest and most diverse set of PTMs --- it
hosts over 60,000 PTMs, fifty times as many as the next largest DL model registry, as well as many types of models and datasets.


\subsection{Deep Learning Trustworthiness}
\label{sec:MachineLearningTrustworthiness}
The trustworthiness (\eg reproducibility, explainability) of DL software grows in importance as DL techniques are deployed in sensitive contexts such as autonomous vehicles~\cite{Injadat2021MLtowardsIntelligentSystems, Marijan2020SWTesting4ML}.
For example, DL traceability is hampered because authors often omit training logs and documentation~\cite{RASHEED2022106043, Rahman2019MLSEinPractice}.
Wing urges the DL community to explore a combination of approaches to achieve trustworthy DL~\cite{Wing2021TrustworthyAI}. 
To improve the trustworthiness of ML systems, prior work recommends considering aspects including provenance, reproducibility, and portability~\cite{Marcal2021Traceability4TrustworthyAI, Floridi2019Rules4TrustworthyAI, Wing2021TrustworthyAI, Rahman2019MLSEinPractice}, as defined in~\cref{table:DLAttributes}.
Some researchers have investigated the performance variances tied to DL frameworks~\cite{park, liu}, which threatens DL reliability. 

Adversarial attacks and defences are also important to DL trustworthiness~\cite{ Akhtar2018AdversarialAttacksonDLinCV, liu2018trojaning}.
Gu \etal proposed the general term \emph{BadNet} for models that perform well on benchmark datasets but poorly on attacker-defined inputs~\cite{Gu2019BadNets}.
Kurita \etal showed that it is possible to construct \emph{BadNets} from weight poisoning attacks by injecting PTM with vulnerabilities that expose backdoors after fine-tuning
~\cite{Kurita2020WeightPoisoningPTM}.
Additionally, Goldblum \etal discussed that it is also possible to attack a model indirectly via malicious labels in its training dataset (data poisoning)~\cite{Goldblum2022DatasetSecurity}.
Wang \etal described an \emph{EvilModel} where a PTM has malware bytes hidden inside its neurons' parameters to be extracted and assembled into malware at run-time~\cite{Wang2021EvilModel}.
These attacks are not all covered by existing malware detection techniques and raise potential risks to DL model registries~\cite{Aslan2020ComprehensiveReviewonMalwareDetectionApproaches}.

\section{Research Questions} 
\label{sec:RQs}

Summarizing the literature:
\new{Much is known about software engineers' practices and challenges in reusing traditional software packages, but little about DL software packages (PTMs).
Reuse and trust are unexamined in DL model registries.} 


We studied the reusability of PTM packages in DL model registries, examining qualitative and quantitative aspects. 
We focused on one DL model registry, \HF, as it is by far the largest registry at present~\cite{2022JiangEmpirical}.
For PTM reuse in the \HF ecosystem, we ask:

\begin{itemize}[leftmargin=25pt, rightmargin=10pt]
    \item[\textbf{RQ1}] \emph{How do engineers select PTMs?}
    \item[\textbf{RQ2}] \emph{What PTM attributes facilitate PTM reuse?} 
     \item[\textbf{RQ3}] \emph{What are the challenges of PTM reuse?}
    

    \item[\textbf{RQ4}] \emph{To what extent are the risks of reusing PTMs mitigated by \HF defenses?}
\end{itemize}

\new{RQ1-2 are focused on current software engineering practice, priming the participants to describe their challenges in RQ3. RQ4 complements this data with quantitative measurements.}

      

\section{Methodology}
\label{sec:Methodology}
To answer our research questions, we used a mixed approach that combined two perspectives~\cite{MixedMethodsResearch}. 
We first explore qualitative insights by interviewing practitioners, then we substantiate our findings with systematic measurements in \HF ecosystem.
The relationship between our questions and methods is shown in \cref{fig:Methods-RQs}.


{
\begin{figure}[ht]
    \includegraphics[width=0.99\linewidth]{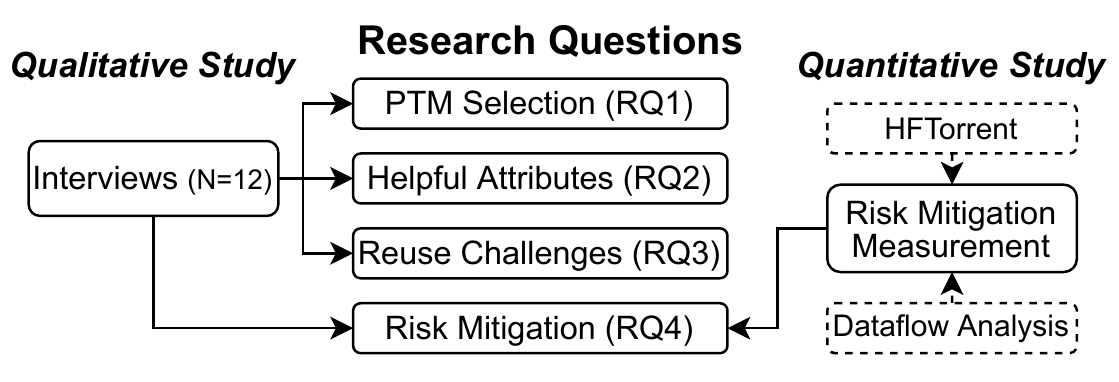}
    \caption{
    Relationship of research questions to methodology.
    }
    \label{fig:Methods-RQs}
\end{figure}
}

\subsection{Qualitative Study: Interviews with PTM Reusers} 
\label{sec:Method-Qual}



\new{Our interview study follows a four-step process modeled on the \emph{framework analysis} methodology~\cite{ritchie2002qualitative, Srivastava2008FrameworkAnalysis}:
}

\myparagraph{(1) Data Familiarization and Framework Identification}
\new{Our initial thematic framework is based on three themes from our literature review (\cref{sec:Background}): model selection, PTM attributes, and PTM trustworthiness.}
For \emph{model selection}, \new{the identified considerations were the PTM reuse issues and factors affecting} the decision-making process~\cite{2022JiangEmpirical, michael2019RegexesareHard}.
For \emph{attributes}, \new{we saw both traditional attributes (\ie popularity, quality, maintenance)~\cite{NPMsAttribute, Abdellatif2020SimplifyingSearchofNPMPackages}, and DL-specific attributes}, viz. provenance, reproducibility, and portability~\cite{Marcal2021Traceability4TrustworthyAI, Wing2021TrustworthyAI, Floridi2019Rules4TrustworthyAI}, shown in the first three columns in \cref{table:DLAttributes}.
For \emph{trustworthiness}, we considered the aspects assumed trustworthy plus possible discrepancies~\cite{2022JiangEmpirical}. 

\myparagraph{(2) Interview Design}
%
%
%
\new{We designed a semi-structured interview protocol with questions that explore the three identified themes of PTM reuse and trust.} 
%
%
\new{We conducted three pilot interviews.
We then revised our framework and interview protocol, adding some PTM attributes and factors and clarifying definitions.}

The final interview protocol took 30--45 minutes.
We compensated interview participants with a \$20 gift card.
The protocol is available in our artifact (\cref{sec:Reproducibility}).

{
\small
\begin{table*}[ht]
\caption{
    Definition and examples of DL-specific attributes, and the relevant factors mentioned by multiple participants.
    }
\label{table:DLAttributes}
\begin{tabular}{
c p{0.3\linewidth} p{0.18\linewidth} p{0.32\linewidth}
}
\toprule
\textbf{Attribute} & \textbf{Definition}    & \textbf{Example} & \textbf{Identified Factors}                  \\
\midrule
Provenance 
& A measure of model lineage or traceability.
& Noting the original paper
& (1) Dataset details (2) Performance table (3) Architecture details (4) Training logs
  \\
\\
Reproducibility 
& The ability of a DL practitioner to produce the same accuracy and latency from a PTM as defined in its paper, source code, or group.
& Providing details of environment configuration
& (1) Hardware specification (2) Training configuration (scripts, hyper-parameters) (3) Demos (4) Documentation (5) Environment image
  \\
\\
Portability  
& The ease with which an engineer can take a PTM and reuse it in another environment, software project, \new{or other application domain}.
& Hardware accelerator information helps engineers know whether a model can run on their devices.
& (1) Hardware specification (2) Latency (3) Quantized model (4) Environment image (5) Framework support (6) Fine-tuning instructions (7) License (8) Cost estimation
\\
\bottomrule
\end{tabular}
\end{table*}
}

\myparagraph{(3) Recruitment}
We recruited users from the \HF ecosystem~\cite{HuggingFaceWeb1}, who presumably have experience in developing and reusing PTMs.
According to the \HF website~\cite{HuggingFaceUserPage} there are \HFNumUsers\ \HF users, \HFNumPROUsers of whom have PRO accounts and \HFNumRegularUsers of whom have regular accounts. 
We sorted the lists of PRO and regular users by the number of models they have contributed to \HF, and contacted the first 50 users of each type.
We interviewed the \interviewNum respondents described in \cref{table: participants}.
This was a (response rate of \interviewRate, of whom \interviewNumPRO had PRO accounts.
Our participants contributed between 4 and 2500 models to \HF.

{
\begin{table}[t]
\centering
\caption{
  Participant demographics.
  Participants
    generally identify as software, ML, or NLP engineers,
    work for small, medium, and large technology companies,
    and
    claim intermediate or expert skill in deep learning (DL) and software engineering (SE).
  Eleven partcipants work with Natural Language Processing (NLP) models, six with Computer Vision (CV) models.
  } 
\label{table: participants}
\begin{tabular}{
    cccccl
}
\toprule
    \textbf{ID} &
    \textbf{Role} &
    \textbf{Org. size} &
    \textbf{DL skill} &
    \textbf{SE skill} &
    \textbf{Domain}
    \\
\midrule
P1                    
    & Tech lead 
    & Small
    & Expert
    & Interm.
    & NLP, CV
\\

P2           
    & Tech lead 
    & Small
    & Interm.
    & Interm.
    & NLP
\\

P3         
    & Engineer 
    & Small
    & Expert
    & Interm.
    & CV
\\
P4               
    & Engineer 
    & Medium
    & Expert
    & Interm.
    & NLP
\\
P5       
    & Tech lead 
    & Large
    & Interm.
    & Interm.
    & NLP, CV
\\
P6                      
    & Engineer 
    & Small
    & Interm.
    & Interm.
    & NLP, CV
\\
P7                      
    & Engineer 
    & Small
    & Interm.
    & Expert
    & NLP 
\\
P8                      
    & Engineer 
    & Small
    & Interm.
    & Interm.
    & NLP
\\
P9                      
    & Data scientist
    & Large
    & Interm.
    & Interm.
    & NLP
\\
P10                      
    & Engineer 
    & Medium
    & Expert
    & Expert
    & NLP, CV
\\
P11                      
    & Engineer 
    & Medium
    & Expert
    & Expert
    & NLP, CV
\\
P12                      
    & Engineer 
    & Small
    & Interm.
    & Interm.
    & NLP 
\\
\bottomrule
\end{tabular}
    \label{tab:symbols}
\end{table}
}

\myparagraph{(4) Analysis}
\new{We transcribed the interview recordings.}
Two researchers performed memoing~\cite{saldana2011fundamentals}, \new{mapping the transcripts to the pre-defined themes.
Each memo had a quote for one of the themes.
Multiple researchers analyzed 4 transcripts and had high agreement on the memos extracted for each theme.
Agreement was because the pre-defined themes had clear definitions, but we did not measure the agreement precisely. A single researcher memoed the remaining 8 transcripts.}

\new{Then we}
\new{organized the memos in a matrix by theme.}
Two researchers used the matrix to develop a thorough understanding of the larger picture.
Then we answered each RQ by our understanding and reference to the matrix.

As part of our analysis, we measured saturation from our interview transcripts \new{by analyzing the number of cumulative unique codes by participant~\cite{Guest2006InterviewDataSaturationandVariability}}.
Saturation was achieved after 7 participants so we did not continue to recruit participants.

\subsection{Quantitative Study: Risk Mitigation Measurement} 
\label{sec:Method-Risk}
Our qualitative findings identified a variety of challenges and risks in PTM reuse.
We measured these risks and mitigations in the \HF ecosystem with the STRIDE
  methodology for threat modeling and risk assessment~\cite{MicrosoftSTRIDE}.\footnote{STRIDE is a mnemonic for Spoofing identity, data Tampering, Repudiation, Information Disclosure, Denial of Service, and Elevation of privilege.}
%
%
%
\new{STRIDE was proposed by Microsoft as a security analysis technique and is widely used~\cite{AWSSTRIDE, MicrosoftSTRIDE, OWASP_STRIDE, shostack2008MicrosoftSTRIDEExperience}.
STRIDE focuses on trust assumptions related to data, making it suitable for PTMs.}

\new{STRIDE is a two-step process. 
First, the system under consideration is modeled using a dataflow diagram, and trust boundaries and the actors involved are identified. 
Second, each boundary is analyzed for the threats of the STRIDE acronym.} 

\begin{figure*}[!ht]
    \centering
    \includegraphics[width=.93\linewidth]{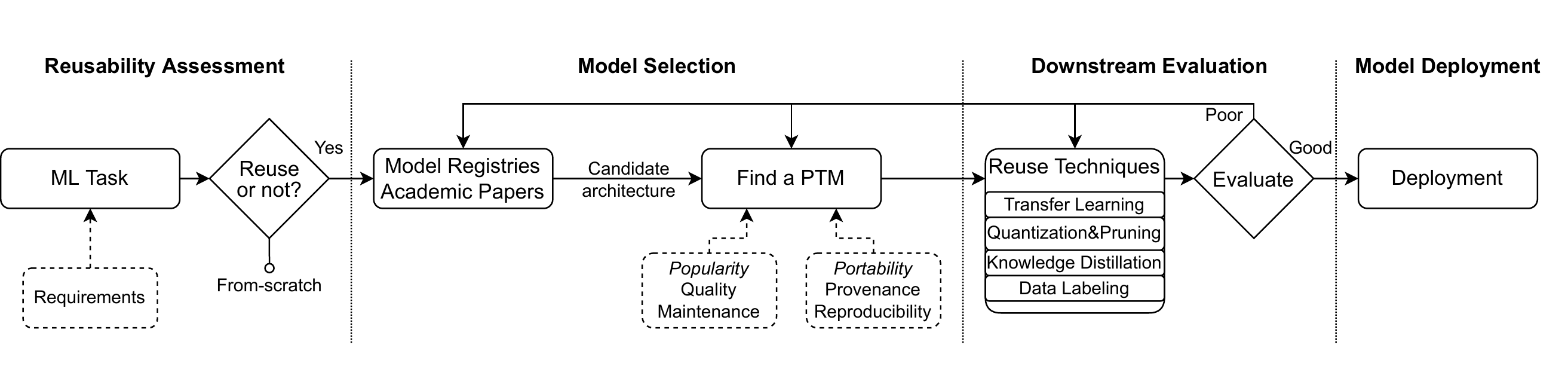}
    \caption{Summarized decision-making progress of PTM reuse. Back edges indicate the possible changes of model selection.
    }
    \label{fig:DecisionMaking}
\end{figure*}


\new{Following the STRIDE methodology, we started by developing a dataflow diagram 
for PTMs on \HF.
Two researchers analyzed \HFs public documentation.
After internal iteration, we settled on one primary trust boundary: user control vs. \HF internal control.
We identified threats and six risk-mitigating features across this boundary.} \new{
Owing to the nature of the available data source (public documentation), we limited our analysis to a subset of the STRIDE threats: Spoofing, Tampering, Repudiation, and Elevation of Privilege.} 
\new{The completeness of our dataflow diagram was ensured by having two researchers review the documentation. 
These same researchers checked the soundness of the model by creating and using models on \HF both as individual accounts and organization contributors.}

\section{RQ1: How do engineers select PTMs?}
\label{sec:RQ1}

\begin{tcolorbox} [width=\linewidth, colback=yellow!30!white, top=1pt, bottom=1pt, left=2pt, right=2pt]
\textbf{Finding 1}: 
The participants share a similar decision-making process (\cref{fig:DecisionMaking}). 
Among the four PTM reuse scenarios in the research literature, our participants reported using only two: transfer learning and quantization techniques.
When reusing, participants find PTMs from DL model registries easier to adopt than PTMs from GitHub projects. 
\end{tcolorbox}

\subsection{Reuse scenarios} \label{sec:ReuseScenarios}

Most interview participants take PTMs from model registries and apply transfer learning techniques to the model.
They either 
\inlinequote{fine-tune an existing PTM} by (optionally) extending architecture and training on a task-specific dataset,
or 
\inlinequote{build a new model on top of the pre-trained one}.
Commonly, they select PTMs from leading technology 
companies (\eg Google, Meta) because \inlinequote{the datasets are carefully cleaned and \new{[the models] are} straightforward to fine-tune}.




The other three reuse scenarios discussed in the research literature (\cref{sec:PreTrainedModelReuse}) were far less common in our interviews.
\emph{P5} described using quantized models. 
No participants described using PTMs for knowledge distillation or for data labeling.

\subsection{Decision-making process}  

To understand how engineers select PTMs, we asked participants to summarize their decision-making processes.
\new{We found similarities between participant responses. 
We followed Michael \etal~\cite{michael2019RegexesareHard} in adapting a general software engineering reuse process~\cite{pressman2005software} to integrate our findings into a unified model (\cref{fig:DecisionMaking}).
}
%
Our model contains 4 stages:
  (1) \emph{Reusability assessment},
  (2) \emph{Model selection},
  (3) \emph{Downstream evaluation},
  and
  (4) \emph{Model deployment}.
We discuss each in turn.

\myparagraph{Reusability Assessment}
Engineers begin the decision-making process with a \emph{reusability assessment}.
Before selecting a model, engineers must identify an ML task and determine if model reuse is appropriate.
An ML task is \new{composed of} 
requirements including model input and output, latency, size, and licensing.
Engineers must then decide if they should reuse a PTM or create a solution from scratch \new{since \inlinequote{PTMs do not work for every use case.}}
Task parameters influence this decision.
For example, three participants (\emph{P8, P9, P11}) reuse PTMs because they \inlinequote{do not have enough computational resources}.
In a similar manner, participants (\emph{P2, P5, P8, P9, P11}) note that DL model registries provide inference APIs to simplify reuse --- PTMs are \inlinequote{easy to use and test}.

\myparagraph{Model Selection}
Once engineers decide to reuse a model, they must select an existing architecture and an associated PTM.
Engineers search for candidate architectures \new{\inlinequote{built for the problem that [they are] trying to solve}},
browsing model registries or relevant papers.
Most study participants prefer to search through model registries.
For example, participants (\emph{P1, P4, P6}) said they can \inlinequote{easily find a model} in model registries due to classification at the domain (\eg computer vision, natural language processing) and task (\eg text generation, image classification) levels.
\emph{P10} noted that standardizing PTM reuse increases model registries' popularity.

Once engineers select a candidate architecture, they must find a \new{particular} PTM \new{to use}.
All study participants, including those who select architectures from papers, prefer PTMs from model registries.
\inlinequote{Ease of use is very important} to engineers that do not think it is worth \inlinequote{spend[ing] much time on trying to understand the script[s] from the GitHub models}.
\emph{P8} noted models from \HF are \inlinequote{plug and play.}
Since multiple PTMs might implement the same architecture, engineers select from among candidates based \new{on} PTM attributes (\cref{sec:RQ3}).
For example, most participants use popularity as a factor to select a PTM because it indicates 
\inlinequote{\new{[community]} trust in the model}.
As another example, multiple participants (\emph{P2, P3, P8, P11}) choose NLP PTMs trained on appropriate datasets (\eg models trained on datasets with the correct language). 


\myparagraph{Downstream Evaluation}
After selecting a PTM, engineers conduct a \emph{downstream evaluation} for their specific task. 
\new{Engineers have the option of assessing more than one candidate PTM in this stage.
They download \inlinequote{a few models,} \inlinequote{finetune them,}  \inlinequote{test them,} then \inlinequote{compare them.}}
When engineers select a candidate PTM, they \new{first} apply reuse techniques to fit the model to their specific application (cf. \cref{sec:PreTrainedModelReuse} and \cref{sec:ReuseScenarios}).
This procedure is not necessarily straightforward because some models \inlinequote{don't really work too well directly, even with their own datasets}.
Furthermore, 11 out of 12 of the participants observed a lack of adequate documentation or discrepancies within existing documentation.

After applying reuse techniques, engineers evaluate the model to see if it is ready for \emph{model deployment}.
\new{When evaluating the trade-off between performance and architecture,}
\emph{P1}, \emph{P8} and \emph{P10} state that these two factors are \inlinequote{tightly relevant to each other} and should be considered in a \inlinequote{fifty-fifty split}.
Some participants prioritize one of these factors over the other.
For example, \emph{P3} and \emph{P6} compare multiple models to maximize task performance.
On the other hand, \emph{P7} will not use a \inlinequote{weird architecture} even if its documented performance is higher.

\myparagraph{Model Deployment}
Finally, engineers deploy their models.
Deployment may depend on model characteristics and deployment environments.
\new{All participants mention} that certain characteristics such as PTM size, robustness, \new{and documentation} significantly impact the deployment of models to \new{other environments (\ie different hardware or software configurations than what is used in development)}.
\new{
Participant (\emph{P8}) describes that the rapid increase in model size makes it \inlinequote{impossible for most \new{[low-resourced teams]} to actually run these models on their systems.}
Participant (\emph{P5}) states that most \inlinequote{models are on PyTorch or TF} and are therefore more difficult to deploy on mobile devices. 
Participant (\emph{P4}) notes that documentation \inlinequote{for running a model on multiple GPUs} is \inlinequote{not clear.}
}

\section{RQ2: What PTM attributes facilitate PTM reuse?}
\label{sec:RQ2}

\begin{tcolorbox} [width=\linewidth, colback=yellow!30!white, top=1pt, bottom=1pt, left=2pt, right=2pt]
\textbf{Finding 2}: 
For \emph{Traditional attributes}, Popularity is most helpful.
Provenance, Reproducibility, and Portability are the three \emph{DL-specific attributes} we should consider.
\end{tcolorbox}


Here we would like to learn about what sort of information is useful to engineers who reuse PTMs.
We asked about two types of attributes here: traditional and DL-specific attributes.

\subsection{Traditional Attributes}
In the interview, we asked about whether the traditional attributes as offered by traditional package registries, such as NPM~\cite{Abdellatif2020SimplifyingSearchofNPMPackages}, are helpful in DL model registries. 

Almost all participants highlighted the importance of \emph{popularity} in DL model registries.
For example, \emph{P12} stated that a PTM with \inlinequote{lots of downloads} means that \inlinequote{it could be a good start point to try}.
\emph{P5} mentioned that \inlinequote{popularity usually goes side by side with maintenance and can indicate the quality}.

Some participants thought that quality and maintenance are also very useful.
\emph{P2} said it is important to \inlinequote{know that it is constantly maintained and does not have many open issues}, and pointed out that good \emph{maintenance} means \inlinequote{if the code is being updated or if you raise a bug, then someone will help you out}. This is highly important because \inlinequote{you are relying on someone else} and \inlinequote{you want to build that trust factor}.

However, some participants think that maintenance and quality are less useful.
Recall from \cref{sec:RQ1} that most reuse scenarios are fine-tuning on new datasets or tasks. 
Provided that the model is fine-tunable, some participants mentioned that maintenance and quality metrics are \inlinequote{less useful in downstream tasks}.
\emph{P12} suggested that maintenance may be less relevant because of the cost of making changes --- \inlinequote{it is really hard to modify the PTM...there were some issues during pre-training} in a large language model, but the model has not been retrained \inlinequote{because it is too large}.




\subsection{DL-specific Attributes}


We added three DL-specific attributes: provenance, reproducibility, and portability. 
The participants provided the relevant factors for each attribute. \cref{table:DLAttributes} also indicates the factors for each attribute which were mentioned by multiple participants.
Most participants mentioned that these three attributes can cover all the aspects they would consider. 

\myparagraph{Provenance}
Some \HF PTMs provide many \emph{provenance} metrics, such as information about original paper, dataset, and architecture.
However, these are not detailed enough to fully address model reuse challenges.
\emph{P3} and \emph{P9} would like to see \inlinequote{visualization of model architecture} and the \inlinequote{explanation of changes} compared to the paper. 
\emph{P2, P4, P6, P8} mentioned the importance of more details of datasets because \inlinequote{different authors process data differently, so it cannot be easily compared}.
\emph{P10 and P12} highlighted the importance of training logs because they would like to see \inlinequote{how the PTM was generated} based on the training checkpoints and scripts.


\myparagraph{Reproducibility}
Reproducibility is the most problematic aspect of PTMs.
The reproducibility issues mainly come from two aspects: (1) the configuration of training (2) the understanding of model.
For the training configuration, the participants tend to care more about the hardware specification (\eg types, memory), training configuration (\eg training scripts, hyper-parameters). As a result, they think environment image would be helpful to help them easily configure the settings and make the model more reproducible.
In terms of understanding of the models, different kinds of \emph{demos} (\eg Notebook demo, Fine-tune demo) and \emph{better documentation} would be helpful.

\myparagraph{Portability}
Different models have different deployment constraints, which makes understanding the portability of PTMs helpful for engineers. 
Similar to \emph{reproducibility}, the portability factors include hardware specification and environment. 
Moreover, for deployment, \emph{latency} and \emph{framework support} aspects are essential. 
For example, the inference time and cost of computational resources could be different in different platforms, as mentioned by \emph{P2} and \emph{P6}. 
This information can help engineers understand the portability of PTMs.
\emph{P3} mentioned that \inlinequote{automate[d] creation of other formats of the model for different hardware} could also be very helpful for the deployment.
The quantized model is also \inlinequote{helpful for continuous deployment and fine-tuning}, as mentioned by \emph{P5}. As a result, he also suggested the development of automated quantization.
Some participants also mentioned the fine-tuning instructions, which help them determine whether a model can be used in specific tasks.
For example, \emph{P12} would like to adapt language models to handle programming languages to improve their software testing, and the fine-tuning instructions can help him on deciding which model they should consider.
\emph{P6} also mentioned that if the model registries can provide a \inlinequote{cost estimation for different servers} (\eg different machine classes on a Cloud service provider). 
Moreover, \emph{P1} and \emph{P3} both said that \inlinequote{licensing should be explicit to industry users}. 

\section{RQ3: What are the challenges of PTM reuse?}
\label{sec:RQ3}

\begin{tcolorbox} [width=\linewidth, colback=yellow!30!white, top=1pt, bottom=1pt, left=2pt, right=2pt]
\textbf{Finding 3}: 
Three common challenges for PTM reuse are missing attributes, discrepancies between claims and actual performances, and model risks (\eg privacy issues and unethical models) (\cref{table:Challenges}).
\end{tcolorbox}

\begin{table*}[ht]
\centering
\caption{
  Challenges associated with PTM reuse.
  Third column shows how many participants (of 12) mentioned the challenge.
  }
\label{table:Challenges}
\begin{tabular}{lp{0.6\linewidth}c}
\toprule
\textbf{Challenge} & \textbf{Description} &\textbf{\# Participants} \\
\midrule
Missing attributes     
& There exist missing details in the model registries.
& 10
\\
Discrepancies          
& The claimed performances (accuracy, latency) are often different from the actual performances.
& 9
\\
Security and privacy risks     
& There could exist malicious models which are harmful for security and privacy aspects.
& 3
\\
Model search
& Sometimes the engineers know the model they want but cannot find it.
& 1
\\
Model application
& It is hard for new users to know the  correct application of a model.
& 1
\\
Model flexibility
& Some of the model architectures are inflexible to change for downstream tasks.  
& 1
\\
\bottomrule
\end{tabular}
\end{table*}


\myparagraph{Missing Attributes}
Missing attributes are identiifed as the most challenging problem. Almost every participant mentioned that there are missing details in the model registries, including datasets, licensing, model details, robustness, and interpretability. The attributes are missing for multiple reasons:
Insufficient documentation is one reason. 
\emph{P5} and \emph{P7} observed \inlinequote{missing details of models} in model registries.
For example, \emph{P1} and \emph{P11} found the \inlinequote{performances of the published models are unclear} in the model registries.
\emph{P8} suggested that the reason for under-documentation in \HF is that PTM authors can upload any model; \HF does not enforce any form of documentation.
Another reason is that the PTM authors occasionally do not measure the robustness and explainability of the models---and model registries do not provide an automated approach to measure such attributes. 


\myparagraph{Discrepancies}
Existing discrepancies are another key challenge mentioned by most participants.
\emph{P7} pointed out that \inlinequote{some of the models are over-promising}.
\emph{P2} indicated another reproducibility issue: the \inlinequote{model names are not named correctly and sometimes the provided scripts are broken}. These discrepancies could result in a waste of time and efforts.
\emph{P5} pointed that another reason for this kind of problem is that training configuration details (\ie hyper-parameters) are hard to find.
\emph{P6} also mentioned that some authors only provide a script instead of providing the actual fine-tuned model and corresponding performances in \HF due to the sensitivity of these results.
\emph{P8} and \emph{P9} indicated that they sometimes follow the provided steps, including the models and datasets---even the hardware configurations---and still could not reproduce the results claimed by the PTM authors.

\myparagraph{Model Risks}
There exist potential risks for PTMs in the model registries, including privacy and ethics aspects.
We discussed in \cref{sec:Background} that prior studies have identified many risks of PTMs.
The participants mentioned both internal and external problems in DL model registries.
Internal risks often involve privacy problems of models and data. 
\emph{P2} mentioned that when using the models from model registries, \inlinequote{the model deployment and data transmission are not in their control}. 
They could not directly deploy the model provided by \HF because it is \inlinequote{unreliable to send} their sensitive dataset by \HF inference APIs.
\emph{P8} mentioned that if a model \inlinequote{is trained with malicious intents. It could have a lot of consequences in the real world}. This indicates the potential risks of a malicious model being uploaded to model registries. 

A PTM can be used for unethical or nefarious purposes.
\emph{P8} gave an example of a chatbot created by training on a racist discussion forum. 
This model \inlinequote{created a huge mayhem} because it was publicly released~\cite{GPT4Chan}. 
\emph{P10} observed that it is hard to know ``\emph{what exactly generated the model because most training logs are missing}'' --- the internal biases are concerning and could potentially make the model a BadNet~\cite{Gu2019BadNets}.




\section{RQ4: To what extent are the risks of reusing PTMs mitigated by \HF defenses?} 
\label{sec:RQ4}

\begin{tcolorbox} [width=\linewidth, colback=yellow!30!white, top=1pt, bottom=1pt, left=2pt, right=2pt]
\textbf{Finding 4}: 
Although \HF offers mitigations for many risks, these mitigations are incomplete or not widely adopted (\cref{tab:mitigation-and-risks}).
Model information can be missing or inaccurate due to the self-reporting nature of model metrics (\cref{fig:ModelClaims}).
These risks make the existence of malicious models possible in the model registries.
\end{tcolorbox}

\label{sec:HuggingFace-Dataflow-Model}

\begin{figure*}[ht]
    \centering   
    \includegraphics[width=0.87\linewidth]{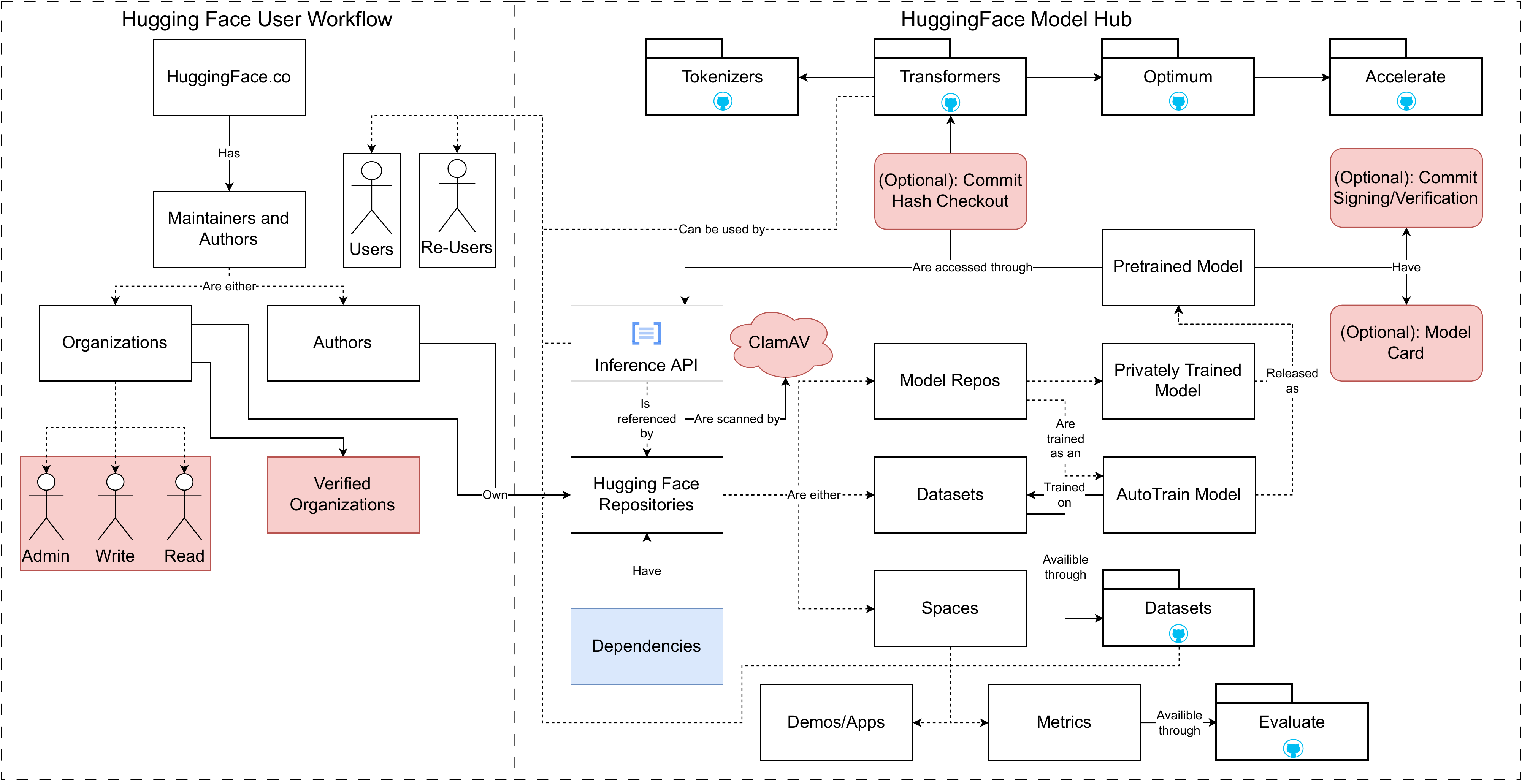}
    \caption{
    Dataflow diagram for PTMs on \HF.
    Security features \new{and dependencies} appear as red \new{and blue blocks}.
    Libraries with the GitHub logo are open source GitHub repositories. 
    \new{The large dashed boxes indicate a trust boundary between users and \HF. Solid connections indicate required paths, and dashed connections represent alternatives.}
    }
    \label{fig:HuggingFace-Dataflow-Diagram}
\end{figure*}


Our interview data identified a range of challenges (\cref{sec:RQ3}).
Incomplete or inaccurate data about PTMs was most common.
Engineers also expressed concern about malicious or unethical models, \eg ``BadNets''~\cite{Gu2019BadNets}.
These findings led us to examine the \HF defenses that mitigate these risks.
We adapted the STRIDE methodology~\cite{MicrosoftSTRIDE} to systematically measure the potential risks in \HF, beginning with an analysis of the dataflow involved in collaborating on \HF. The identified risks are shown in \cref{tab:mitigation-and-risks}.


\subsection{\HF PTM Dataflow Model}
Based on our analysis of \HFs Hub and client libraries documentation, we made a dataflow diagram as shown in \cref{fig:HuggingFace-Dataflow-Diagram}, to represent the how models are created and shared.
This allows us to visualize the dataflow from model contributors to users, which involves the developing, releasing, and accessing of PTMs on \HF.

The common unit of reuse on \HF is the \emph{repository}, classified
  into
  \emph{datasets} (input/output data for supervised or unsupervised learning)
  and
  \emph{models} (PTM architecture, weights, and configuration, cf.~\cref{fig:packages}). 



All \HF repositories have automated and manual \new{risk mitigations to limit} the spread of malware or malicious models.
These include organization verification, user permission models, commit hash checkout, and model cards (\cref{fig:HuggingFace-Dataflow-Diagram}).
\new{Additionally, some PTMs depend on other datasets or PTMs which can expose themselves to outside risks.}

\subsection{Risk Analysis}
\label{sec:riskAnalysis}

\HF implements six risk mitigations.
These are organization verification, \new{PTM documentation, GPG commit signing, a user permission model,  automatic malware scanning (ClamAV), and utilizing a commit hash to checkout a model.}
While commit hash checkouts are not easily measured (performed by users on their own machines), we measured the use and scope of the others within the \HF ecosystem.
We detail our results on \emph{organization verification}, \emph{PTM documentation}, \new{\emph{user permission model}}, and \emph{GPG commit signing}.
\new{Following the risk analysis of traditional package registries~\cite{Zimmermann2019SecurityThreatsinNPMEcosystem}}, we also examine the potential impact of 
\emph{dependencies}.
%
The result for \new{ClamAV was} uninteresting \new{as no (zero) PTM packages were flagged by ClamAV.} 
{
\renewcommand{\arraystretch}{1.5}
\begin{table}
    \centering
    \caption{
    \new{Mitigation and the corresponding STRIDE risks.}
    }
    \begin{tabular}{p{0.22\linewidth}p{0.25\linewidth}p{0.38\linewidth}}
      \toprule
        \textbf{Mitigation} & \textbf{STRIDE risks} & \textbf{Details} \\
         \toprule
        Organization \newline verification  & Spoofing & Low adoption (3.19\%) \\
        Dependencies & Tampering & High dependency (one dataset has 1,673 downstream PTMs) \\
        PTM Documentation & Repudiation & Less than 0.1\% provide machine-parseable claims\\
        GPG commit \newline signing & Spoofing, Tampering, Repudiation & Low adoption (0.21\%)\\
        User permission model & Elevation of privileges  & ``Organizations'' violate the principle of least privilege\\
        ClamAV & N/A & Zero packages were flagged \\
        \bottomrule
    \end{tabular}
    \label{tab:mitigation-and-risks}
\end{table}
}


\myparagraph{Organization Verification}
\HF allows an organization to increase trust by verifying its identity, demonstrating to \HF that an organization controls an associated web domain.
We counted the number of verified organizations via web scraping.
Out of \NOrganizations organizations, only \NVerifedOrgs (\PercentOfVerifiedOrgs) were verified.
With such a small percentage of verified organizations, users cannot determine the legitimacy of unverified organizations.
The low adoption rate raises the risk of \emph{Spoofing}: malicious users may masquerade as real organizations, similar to typosquatting~\cite{Zimmermann2019SecurityThreatsinNPMEcosystem, tschacher2016typosquatting}.

This lack of adoption is concerning because organizational reputation was cited as a factor under the Provenance attribute (\cref{table:DLAttributes}).
The risk of such squatting attacks can be greater for PTM packages than for traditional ones, because new niches in the ecosystem accompany every new state-of-the-art model.
A malicious actor could identify missing PTMs in the cross product of (architecture, dataset) and provide \emph{EvilModels}~\cite{Wang2021EvilModel} in that niche, pretending to be a legitimate organization.

\myparagraph{Dependencies}
%
The dependencies of PTMs pose potential risks.
Malicious models can be injected directly via data manipulation~\cite{Goldblum2022DatasetSecurity}, or indirectly via weight poisoning~\cite{Kurita2020WeightPoisoningPTM}.
Models released through \HF may be vulnerable through their dependencies on model architectures, the associated weights for those architectures, and datasets.
\emph{P5}, \emph{P8}, \emph{P10} stated that they fine-tune PTMs on a daily basis, implying that an attack could have a rapid impact.


\cref{fig:NumberOfModelsPerDataset} shows that the \code{Universal Dependencies} dataset~\cite{nivre2016universal} is the most popular dataset on \HF, with 6,834 models depending upon it.
The distribution of models that depend on a particular architecture is similar to \cref{fig:NumberOfModelsPerDataset}.
We found that the \code{BERT}~\cite{devlin2018bert} architecture is the most popular architecture on \HF presently, with 10,247 models depending upon it.
\HF models have the potential to be trained off of multiple datasets as well.
Our analysis of the existing \HF models shows that many models depend on the works of others that can be maliciously tampered with.
In tampering with these dependencies, \emph{BadNets}~\cite{Gu2019BadNets} and unethical models~\cite{GPT4Chan} can be created, which could affect downstream PTMs.

\myparagraph{PTM Documentation}
\cref{fig:ModelClaims} shows the distribution of missing documentation in \HF.
The highest proportion of models that make performance claims in machine-parseable documentation (YAML) was from the \code{token-classification} task, with 17\% of models meeting the criteria.
We found that 26,192 models belong to various tasks (represented by \code{other}) where \new{only 12 (0.05\%) PTMs} provide machine-parseable claims about the PTM.
\new{Due to the sparse usage of performance claim reporting, there \new{can be potential risks} of Repudiation: model performance can be obscured, misreported, or misleading.}

Some models report their performance in plain texts or tables, and are hard to identify by the users.
It is also common for documentation to omit any performance claims or to refer the user to read an associated research paper for more information about performance, without assurance that this is the same model tested in the research paper. 
Some popular PTMs are poorly documented as well.
The language model \code{SpanBERT/spanbert-large-cased} is the 9th most downloaded PTM on \HF and receives 6.9 million monthly downloads, yet has no model card.   
\cref{fig:ModelClaims} supports that \emph{missing attributes} is a real challenge existing in the \HF DL model registries. 
The lack of transparency reduces the trustworthiness of the models and increase the potential risks of malicious models.

{
\begin{figure}[ht]
    \includegraphics[width=0.92\linewidth]{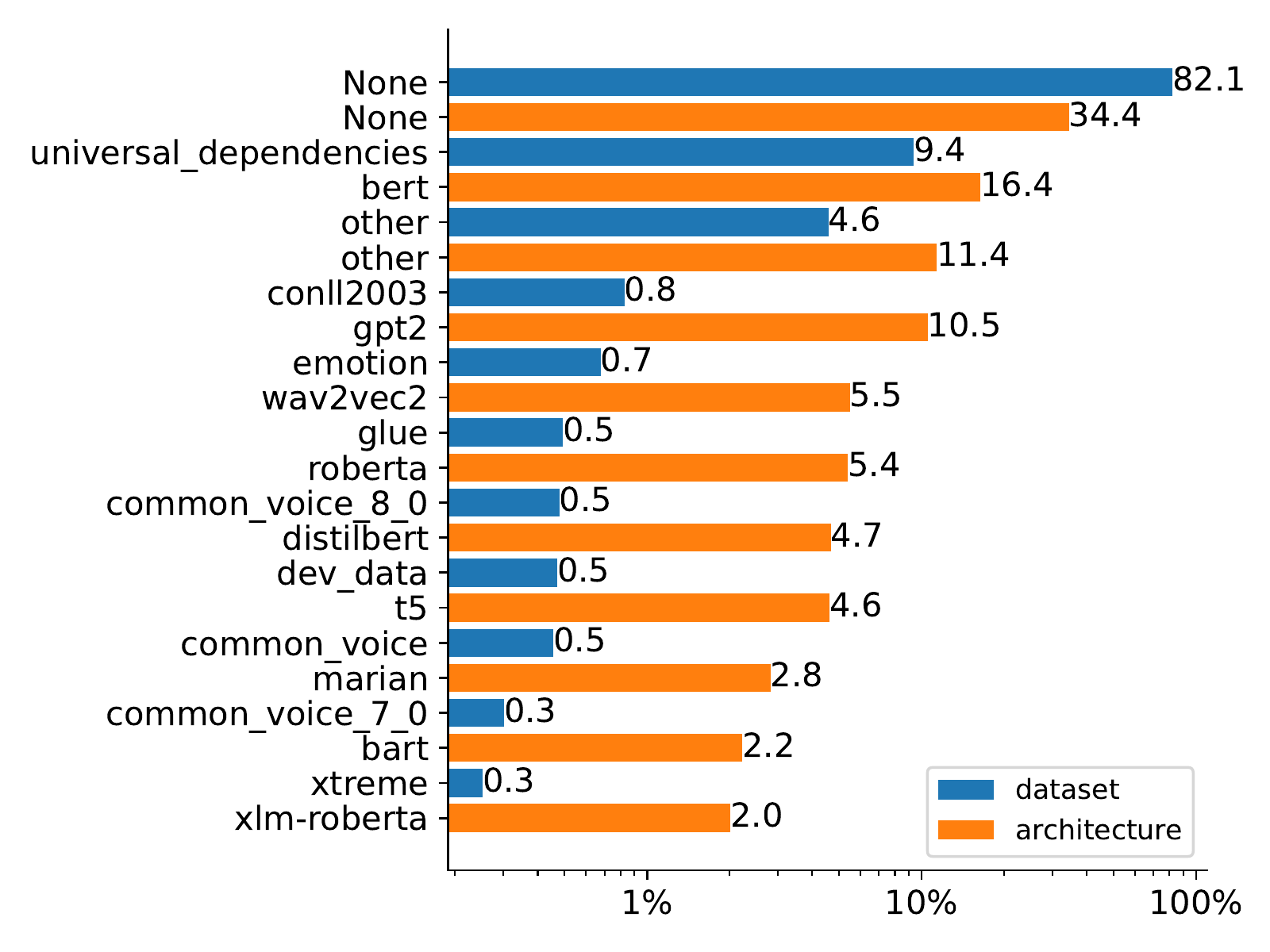}
    \caption{
    Number of models trained on a specific dataset \new{(blue) and trained with a specific model architecture (orange).
    1,207 models were trained on the most popular dataset. 1,673 models depend on the most popular architecture.}
        }
    \label{fig:NumberOfModelsPerDataset}
\end{figure}
}

\myparagraph{GPG Commit Signing}
\label{sec:GPGCommitSigning}
\HF provides a verification tag for commits that have been signed with a GPG key.
By signing with a GPG key, users are providing verification that they have signed their commits, and not as someone masquerading as them.
This feature allows users to accurately trace back the changes to the PTM packages.
Using the HFTorrent dataset (\cref{sec:HFTorrent}), we analyzed the usage of commit signing within \HF model repositories.
Out of \PTMDatasetNPackages model repositories, we found \PTMDatasetNReposWithSignedCommits (\PTMDatasetPercentOfSignedCommits) repositories within the dataset implemented signed commits.
Additionally, only 2 verified organizations have at least one repository 
with signed commits.
This indicates that potential attackers can be contributing malicious code, implementing a \emph{BadNet}~\cite{Gu2019BadNets} or \emph{EvilModel}~\cite{Wang2021EvilModel} under a pseudonym, or manipulating the \code{git} commit history to hide malicious activity.

\new{
The limited usage of GPG commit signing exposes models hosted 
on \HF to \new{\textit{Spoofing}, \textit{Tampering}, and \textit{Repudiation} risks.}}
\new{
First, unsigned \HF models are vulnerable to \textit{Spoofing} since attackers could make commits under the alias of a legitimate maintainer.
Second, these models face the risk of \textit{Tampering} because attackers could contribute malicious code or edit commit history.
Finally, unsigned models could also risk \textit{repudiation} since a lack of verified commit history allows an individual to deny actions within a repository.
}

\myparagraph{User Permission Model}
\HF has a standard approach of users and organizations (\cref{fig:HuggingFace-Dataflow-Diagram}). 
One shortcoming of the permission model is that \HF organizations violate the principle of \textit{least privilege}\new{~\cite{saltzer_protection_1975}}:
an organization member with \textit{Write} privilege can modify any PTM owned by the organization.
\new{Therefore, an attacker could contribute malicious code, thereby raising the risk of \emph{Elevation of Privileges}.}

{
\begin{figure}[t]
    \includegraphics[width=0.9\linewidth]{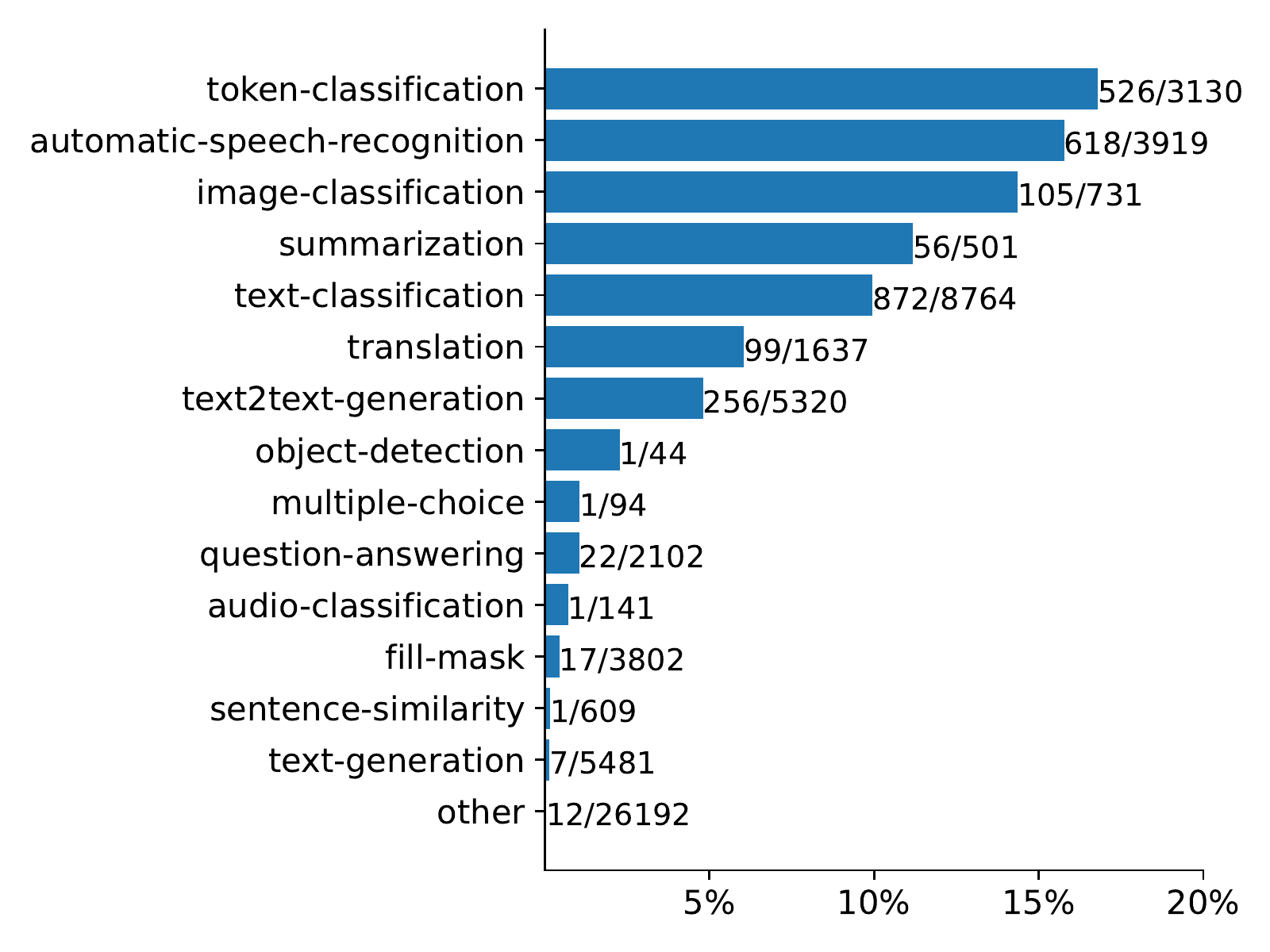}
    \caption{
    The proportion of models with standardized claims (\HFs YAML format) for each type of PTM.
    }
    \label{fig:ModelClaims}
\end{figure}
}

\section{The HFTorrent Dataset} 
\label{sec:HFTorrent}

Our analysis in \cref{sec:RQ4} relied in part on measurements of the PTM packages in \HF.
To reduce the impact on the \HF service during our measurements, we took a snapshot of \PTMDatasetNPackages PTM packages in the \HF registry. 
This snapshot, the \emph{HFTorrent Dataset}, is included in the artifact accompanying this paper.
An improved version of this dataset is now available~\cite{Jiang2022PTMTorrentFigshare}.
%
We hope these data facilitate further research on PTM packages, similar to the impact of the GHTorrent~\cite{Gousios2012GHTorrent} and SOTorrent~\cite{Baltes2018SOTorrent} datasets.

\myparagraph{Creation process}
We initiated a copy of all PTM packages in the \HF registry.
Copies were taken between August 15th and 16th, with rate limiting to avoid abuse of the \HF registry.
\PTMDatasetFailedPackages (\PTMDatasetFailedPercentage) of the copies failed, caused, we believe, by concurrent changes in package names.

\myparagraph{Dataset contents}
The HFTorrent dataset contains the repository histories of \PTMDatasetNPackages of the PTM packages available on \HF as of August 2022.
They are provided as bare git clones to reduce space, resulting in a compressed footprint of $\sim$20\ GB. 
Each PTM package can be reconstructed to its most recent version, including the model card, architecture, weights, and other elements provided by the maintainers (cf. \cref{fig:packages}).
Further information is available in our artifact.



\section{Discussion and Implications}
\label{sec:discussion}

\subsection{Integrating the findings}

\new{Our qualitative and quantitative studies provide a deeper understanding of the practices and challenges for DL model registries.
In \cref{sec:RQ1} we obtained a general reuse process (\cref{fig:DecisionMaking}) which is complemented by the specific details for \HF (\cref{fig:HuggingFace-Dataflow-Diagram}). In \cref{sec:RQ2} we studied how the theoretical attributes of reproducibility and provenance can affect the decision-making process of PTM reuse (\cref{fig:DecisionMaking}). These attributes are partially operationalized in \HF by aspects measured in \cref{sec:RQ4}: organizations and verification, PTM documentation, GPG commit signing, and dependencies. For example, our results in \cref{sec:RQ4} imply that although PTM reusers value the provenance of models, this provenance is actually untrustworthy in \HF due to the low adoption of verified organizations and commit signing.}
\new{In \cref{sec:RQ4} we measured risks based on the identified challenges from \cref{sec:RQ3}. For example, in \cref{sec:RQ3} we qualitatively learned that documentation may be missing or have discrepancies. In \cref{sec:RQ4} we quantitatively measured that model documentation is missing or inadequate for 80\% of PTMs.}


\subsection{Comparison to traditional package registries}

\begin{table}
    \centering
    \caption{
    \new{Main differences between DL model registries and traditional (\textit{trad.}) package registries.}
    }
    \begin{tabular}{
    p{0.2\linewidth}p{0.6\linewidth}
    }
          \textbf{Aspect} & \textbf{Observed difference from trad. registries} \\
         \toprule
        Decision-making process  & Selecting PTMs needs a more complicated assessment of requirements and trade-offs, and several iterations of downstream evaluation.  \\
        \midrule
        Attributes & PTM reusers have to consider additional DL-specific attributes.  \\
        \midrule
        Potential risk & There exist traditional risks and additional PTM risks in model registries. \\
        \bottomrule
    \end{tabular}
    \label{tab:compare_PTM_traditional}
\end{table}

Our qualitative analysis in \cref{sec:RQ1}---\cref{sec:RQ4} sheds light on the differences between model registries and traditional registries, in terms of decision-making, attributes, \new{and potential risks}.

\myparagraph{Decision-making}
\new{Our results indicate that the decision-making process of PTMs (\cref{fig:DecisionMaking}) is more complex than traditional packages~\cite{Abdalkareem2017TrivialPackages, Bogart2015HowEcosystemDevelopersReasonabouttheStabilityofDependencies, Jadhav2011Framework4EvaluationaandSelectionofSWPackages}, both in terms of the assessment and evaluation.}
Traditional software package reuse is integrated throughout the software development process~\cite{Anasuodei2021SWReusabilityApproachesandChallenges} to improve productivity~\cite{Abdalkareem2017TrivialPackages}. 
Our result shows that DL engineers behave similarly with PTMs.
\new{However, PTM reusers have to perform more complicated assessment and evaluation during decision-making process.}
\new{PTMs are hard to evaluate and compare with others, as indicated by \textit{P3, P6, P8, P10}}.
Moreover, \cref{fig:DecisionMaking} involves three back edges (loops between selection and assessment), while the decision-making process for the traditional software reuse \new{reportedly involves fewer iterations~\cite{Jadhav2011Framework4EvaluationaandSelectionofSWPackages, Jadhav2009EvaluatingSelectingSWPackages}}.
Moreover, we can see in the selection stage of \cref{fig:DecisionMaking}, different factors are considered, including the dataset availability and cost. Traditional software reuser tend to consider more about the ease of use, and such detailed information are less needed by PTM reusers.
Wang \etal suggest that
developers should use the usage statistics to guide the evolution~\cite{Wang2020EmpiricalStudyofThirdPartyLibsinJavaProjects}.
Similarly, PTM providers should consider the practices and challenges based on our qualitative data, and utilize it as a guidance to improve the PTMs in DL model registries.

\myparagraph{Attributes}
Our interview data indicated the significance of traditional attributes (\ie popularity, quality, and maintenance) and the requirements for DL-specific attributes. 
For \emph{Traditional attributes}, \emph{Popularity} is most helpful for PTM reuse, while \emph{Quality} and \emph{Maintenance} are less useful compared to traditional packages~\new{(\cref{sec:RQ2})}. 
This differs from traditional software reusers, who consider quality and maintenance of packages as important as popularity when selecting and reusing the packages~\cite{Abdellatif2020SimplifyingSearchofNPMPackages, Mujahid2022CharacteristicsofHighlySlectedPackages, Zerouali2019DiversityofSWPackagePopularityMetrics}.
This difference comes from the expensive cost and data-driven nature in PTMs. It is hard for PTM users to directly obtain and employ an existing model. In contrast with taking a package from NPM or PyPI and directly reuse it in the codes, PTM reuse requires a deeper understanding and knowledge of how a model works. However, \inlinequote{there are not quite reliable methods to measure the explainability of PTMs yet}, as indicated by \emph{P10}. Therefore, to better reuse the PTMs, the engineers need more information from the model registries, which result in the requirements for DL-specific attributes. 


\myparagraph{Potential risks}
\new{The dependencies, maintainers, and reported issues can help analyze the security risks in software registries~\cite{Zimmermann2019SecurityThreatsinNPMEcosystem}.
Prior work indicates that developers should manage the upstream dependencies and minimize the impact on downstream tasks~\cite{Bogart2015HowEcosystemDevelopersReasonabouttheStabilityofDependencies}.
Recently, Jiang \etal empirically studied the maintainers' reach of model registries~\cite{2022JiangEmpirical}. However, the results here suggest that there are different risks within model registries, such as Spoofing, Tampering, Repudiation, and Elevation of privileges (\cref{tab:mitigation-and-risks}).
These risks can have varying degrees of impacts on the reusability of PTMs:
\new{The unaware discrepancies \inlinequote{do play a huge role} (\textit{P8}, \textit{P11}, \textit{P12}). They may not only \inlinequote{hinder developer progress}, but also change the accuracy and robustness of downstream PTMs~\cite{Goldblum2022DatasetSecurity}. Moreover, our STRIDE analysis (\cref{sec:RQ4}) indicates multiple risks and the potential to introduce vulnerabilities~\cite{Gu2019BadNets, Kurita2020WeightPoisoningPTM}.}}

\subsection{Implications}


Compared to well-studied traditional package registries (\eg NPM, PyPi), the use and study of DL model registries is still in its infancy. 
Our empirical study of \HF model registry informs future directions on 
model audit, infrastructure, PTM standardization, and adversarial attack detection.


\myparagraph{Model audit}
Our results in \cref{sec:RQ2} and \cref{sec:RQ3} shows that one major challenge of PTM reuse is the missing attributes in model registries. The most important attribute is the performance of PTMs which would significantly affect the reusability of models. Though recently \HF released their automated validation tool~\cite{HFAutoEvaluator}, it is still not able to satisfy the requirement of engineers.
\emph{P8} stated that \inlinequote{Robustness and Explainabiltiy are very key factors to consider when you are deploying ML models in the real world.}
which are important for the eventual goal of model transparency~\cite{JRC119336}.

The proposed DL-specific attributes should also be measured and provided. Our study indicates the importance of these attributes (\cref{sec:RQ2}) and identify the corresponding factors. 
We inform researchers on developing \new{formulas} and automated tools to automatically calculate the score of each attribute and integrate them into the model registries, similar to the \emph{pqm} scores (\ie popularity, quality, and maintenance) in~\cite{NPM}.

We envision that future directions should consist of large-scale measurements of PTMs or of encouraging model registries to change their PTM release requirements so that it would be easier for users to audit models by themselves.

\myparagraph{Infrastructure}
The infrastructure of model registries can be improved from different aspects.
\emph{P2} mentioned that the badge mechanism would be helpful for communicating the missing attributes (\eg reproducibility), similar to continuous integration badges provided by GitHub~\cite{GitHUbBadge}.
\emph{P6} mentioned a unified fine-tuning process could also assist engineers.
Moreover, multiple participants highlighted the importance of tools for automatically creating quantized models or converting models into different formats. These tools could improve PTM portability and thus support model deployment.

Some interview participants (\emph{P6, P9}) mentioned that they select PTMs first based on their experience, and then based on evaluation metrics.
\emph{P10} mentioned that he reuses PTMs to understand the \inlinequote{the generalizing behavior of fine-tuned models}.
This envisions development of PTM recommendation systems, which can sort the models by scores calculated by model performance in downstream tasks, or predicted by another ML model. A similar system can help to reduce the work for ML engineers and could be integrated as part of the AutoML pipeline for PTM reuse~\cite{He2021AutoMLSurvey}.

\myparagraph{PTM Standardization}
Our results on \cref{sec:RQ2} shows that,
the model registries should include the training logs and corresponding checkpoints. This information can help reusers better understand the PTM and improve the provenance and reproducibility.
As associated research opportunities, such artifacts can be costly to create and store, and it is unclear how engineers can best apply them.

Beyond model provenance, another opportunity for standardization is in the model format itself.
\emph{P5} said that \inlinequote{many PTMs are on PyTorch or TensorFlow} but they would like to use ONNX format models which could make the deployment easier for them. 
However,
due to the rapid appearance of new operators~\cite{mmdnn}, ONNX could not support all of them, especially for state-of-the-art models~\cite{ONNXBlog}.
Knowing the compatibility of PTMs in model registries with standardized formats such as ONNX would also engineers make better decisions and save their time.
We suggest future work examining development challenges in the ONNX framework.

\myparagraph{Adversarial attack detection}
One of the major challenges for PTM reuse are adversarial attacks, as shown in \cref{sec:RQ1}. 
Attacks can be harmful to both the PTM reusers and the downstream application users. 
Although \HF employs ClamAV~\cite{ClamAV} for malware scanning, this only detects traditional attacks, not new attacks such as \emph{BadNets}~\cite{Gu2019BadNets}.
As a result, we suggest future studies working on automated detection of toxic models and poisoned datasets.
Integrating these detectors into model registries can largely improve their trustworthiness.

\section{Threats to Validity}

\vspace{-0.05cm}
\myparagraph{Internal Threats}
Our choice of research methods potentially threatens the validity of our investigation of PTM reuse practices and challenges (RQ1-3).
Our results here are derived solely from interview data but not generalized via a survey instrument. 
As a mitigation, our measurements of the \HF ecosystem (RQ4) substantiate many of the interview participants' concerns.
In addition, we note that the participants of greatest interest are those who also make the greatest use of PTMs, \ie presumably those with PRO accounts on \HF.
\interviewNumPRO of our \interviewNum participants have PRO accounts, representing 1\% of the PRO user population.

Another internal threat is the reliability of our framework analysis on the interview data.
Our framework analysis might be biased by our understanding and transfer of concepts from traditional software to PTMs. 
This conceptual framework helps us tease out similarities and differences in the PTM context, although we 
recognize that our interviews might reflect our biases and perspectives and therefore bias participants to a certain way of thinking.
To mitigate bias, we asked if the participant had anything to add in terms of each theme in our framework throughout the interview, and some did so.

\myparagraph{External Threats}
Our study examines only one DL model registry, \HF.
We note, however, that examining a single package registry, \eg NPM, is fairly common in the literature.
For PTMs, focusing on \HF is sensible, since it is the only open DL model registry and has an order of magnitude more models than other registries.
Our results may not generalize to other DL model registries, but given the relative importance and growing influence of \HF it is unclear whether this is a concern.

\new{Another external threat is the saturation of our interview study because of the interview study’s sampling approach (one PTM registry) and size (12 participants). Within our sample, we saw a high degree of agreement. \new{The saturation of our interview study was achieved after 7 participants (\cref{sec:Method-Qual})}}

\new{
ML researchers identified many uses of PTMs (\cref{sec:PreTrainedModelReuse}), but our participants only employed a subset related to fine-tuning: transfer learning, quantization and pruning. This may pose a threat to external validity.
Our random sampling approach may bias us towards high-probability use cases. One interpretation of our data is that fine-tuning is a popular approach in practice, which would motivate greater study of PTM fine-tuning relative to more theoretical applications. 
}


\section{Conclusion}
We conducted the first empirical study of PTM reuse in the \HF DL model registries.
Based on interviews with \interviewNum practitioners, we defined the decision-making workflow for PTM reuse, and identified three challenges, including missing attributes, discrepancies, and model risks.
To substantiate our qualitative data, we further investigated into useful attributes and potential risks in the \HF ecosystem.
We unveiled risky engineering practices in the \HF ecosystem, particularly a lack of signatures in the PTM supply chain.
Our empirical data motivates future research on PTM audit, automated PTM attribute measurements, improved infrastructure for PTM reuse, and PTM standardization.


\section*{Reproducibility and Research Ethics} \label{sec:Reproducibility}

Our artifact is available at \url{https://doi.org/10.5281/zenodo.7555469}.
Within it, we provide
  the anonymized interview data used to answer RQ1-3,
  the HFTorrent dataset used to answer RQ4,
  and software for the measurements described in RQ4.
Human subjects work was approved by institutional IRB. 


\section*{Acknowledgments}
The authors thank the reviewers; and A. Grigorescu, D. Montes, A. Indarapu, and A. Tewari for their input.
This work was supported by Google and Cisco and by NSF awards \#2107230, \#2229703, \#2107020, and \#2104319.

\raggedbottom
\pagebreak
\balance

\bibliographystyle{refs/IEEEtran}
\bibliography{refs/WenxinZotero, refs/Reference, refs/DualityLab}

%
\raggedbottom
\pagebreak
\balance

\end{document}